\documentclass[%
preprint,
amsmath,amssymb,
]{revtex4-1}

\usepackage{graphicx}
\usepackage{bm}
\usepackage[whole]{bxcjkjatype}
\usepackage[dvipsnames]{xcolor}
\usepackage{url}
\usepackage{hyperref}
\usepackage[caption=false]{subfig}

\usepackage{mathtools} 
\usepackage{fix-cm}
\usepackage{amsmath, amsfonts, bm}

\def\eqref#1{~(\ref{#1})}

\def\1{\bm{1}}

\DeclareMathAlphabet{\mathsfit}{\encodingdefault}{\sfdefault}{m}{sl}
\SetMathAlphabet{\mathsfit}{bold}{\encodingdefault}{\sfdefault}{bx}{n}

\def\gF{{\mathcal{F}}}

\def\gL{{\mathcal{L}}}

\def\gN{{\mathcal{N}}}
\def\gO{{\mathcal{O}}}

\def\gZ{{\mathcal{Z}}}

\newcommand{\pdata}{p_{\rm{data}}}

\newcommand{\E}{\mathbb{E}}

\newcommand{\R}{\mathbb{R}}

\newcommand{\KL}{D_{\mathrm{KL}}}

\DeclareMathOperator{\Tr}{Tr}

\newcommand{\Dn}{D_n}
\newcommand{\gc}{{\rm GC}}
\newcommand{\Ec}{\tilde{\E}_{r,\gc}^{\beta,n}}
\newcommand{\phiz}{\phi_{\backslash z}}

\DeclareRobustCommand{\btheta}[0]{\bm{\theta}}

\newcommand{\eqrefflat}[1]{\textup{(\ref{#1})}}

\begin{document}

\title{Variational Gaussian Approximation in Replica Analysis of Parametric Models}

\author{Takashi Takahashi}
  \email{takashi-takahashi@g.ecc.u-tokyo.ac.jp}
\affiliation{
  Institute for Physics of Intelligence, The University of Tokyo
}
\affiliation{
  RIKEN center for AIP
}

\date{\today}

\begin{abstract}
We revisit the replica method for analyzing inference and learning in parametric models, considering situations where the data-generating distribution is unknown or analytically intractable. Instead of assuming idealized distributions to carry out quenched averages analytically, we use a variational Gaussian approximation for the replicated system in grand canonical formalism in which the data average can be deferred and replaced by empirical averages, leading to stationarity conditions that adaptively determine the parameters of the trial Hamiltonian for each dataset. This approach clarifies how fluctuations affect information extraction and connects directly with the results of mathematical statistics or learning theory such as information criteria. As a concrete application, we analyze linear regression and derive learning curves. This includes cases with real-world datasets, where exact replica calculations are not feasible.
\end{abstract}

\maketitle

\section{Introduction}
\label{section: introduction}
The statistical mechanics analysis of learning and inference from examples has a long history, starting with the seminal work of Sompolinsky, Tishby, and Seung in the early 1990s \cite{PhysRevLett.65.1683,PhysRevA.45.6056}. Since then, such approaches have become a standard topic in the statistical mechanics of disordered systems \cite{engel2001statistical,10.1093/acprof:oso/9780198509417.001.0001}.  One of the central analytical tools in this line of research is the replica method \cite{doi:10.1142/0271,doi:10.1142/13341,MAL-105}. It provides a powerful method to evaluate the average behavior of statistical quantities such as free energy and prediction error. Unlike classical learning theory, which often yields only bounds for those quntities \cite{Shalev-Shwartz_Ben-David_2014}, the replica method can give sharp predictions. 

Despite its strength, the replica method has a major limitation. To take quenched averages analytically, one usually has to assume highly idealized data-generating distributions. This simplification is required even in problems as basic as the standard linear regression. Hence, the replica method has been regarded as a mathematical technique for guessing exact solutions under highly restricted conditions, rather than a tool for predicting universal structures of learning and inference, or for describing the behavior of them in realistic data, although some exceptions exist \cite{NIPS2000_ed519dac, NIPS2001_26f5bd4a,malzahn2002statistical, malzahn2003learning,Malzahn_2005,10.21468/SciPostPhys.4.6.040, NEURIPS2021_9704a4fc, yoshino2020complex,yoshino2023spatially}.

A particularly notable line of work in the context of learning and inference was provided by the series of studies by Malzahn and Opper \cite{NIPS2000_ed519dac, NIPS2001_26f5bd4a,malzahn2002statistical, malzahn2003learning,Malzahn_2005}. By combining the replica method with variational approximations, they successfully obtained an approximate solution to learning curves for Gaussian process regression and hard-margin support vector machines without assuming a concrete data distributions. The use of the variational approximation simultaneously offered three important advantages. They relax the dependence on overly simplified data distributions, and it allowed a systematic way of giving approximate predictions even when exact replica calculations were not trivial. Moreover, it does not necessarily require taking the thermodynamic limit.  These advantages may be useful in the analysis of learning with modern complex models.

However, their analysis was mainly restricted to non-parametric models, and did not address parameter-space formulations that do not necessarily reduce to Gaussian process analyses, which may be more directly relevant to parametric models such as modern neural networks. In this work, we investigate the use of variational Gaussian approximation in replica analysis of parametric models. As a first step, we represent general asymptotic properties as well as application to linear regression. While most of the resulting formulae are already known from mathematical statistics and learning theory, this study represents a methodological step toward relaxing the restricted assumptions of conventional replica analysis of parametric models and may broaden their applicability to a wider class of learning and inference problems.

The remainder of the paper is organized as follows. In Section \ref{section: setup}, we describe the setup of our analysis, focusing on independent and identically distributed (i.i.d.) data and parameter inference based on the Boltzmann distribution, also known as the Gibbs posterior \cite{catoni2007pac, 0a063179-9e5a-36f4-b9fb-7a23e4dd9d21}. Section \ref{section: replica method} introduces the framework of replica method for learning and inference. In Section \ref{section: variational gaussian}, we formulate the variational Gaussian approximation. Section \ref{section: general results} discusses general asymptotic properties. In Section \ref{section: linear models}, we apply the method to linear regression as a concrete example. Finally, Section \ref{section: summary} summarizes the results and outlines possible directions for future research.

\section{Setup}
\label{section: setup}
We consider a dataset $\Dn = \{z_i\}_{i=1}^n$ of $n$ i.i.d. samples drawn from an unknown distribution $\pdata$ defined on a sample space $\gZ$. 
Each data point $z_i$ represents a generic observation. 
In unsupervised learning it can be a feature vector $x_i \in \mathbb{R}^d$, 
and in supervised learning it corresponds to an input--output pair $(x_i, y_i)$.

In order to describe learning and inference with a parametric model, we introduce a Boltzmann distribution, also known as the Gibbs posterior \cite{catoni2007pac, 0a063179-9e5a-36f4-b9fb-7a23e4dd9d21}, on the parameter space $\Theta$:
\begin{equation}
  p^{\beta}(\theta \mid \Dn)
  = \frac{1}{Z^\beta(\Dn)} e^{-\beta \, \gL(\theta;\Dn)},
  \label{eq: boltzmann}
\end{equation}
where
\begin{equation}
  \gL(\theta;\Dn) = \sum_{i=1}^n l(\theta;z_i) + \frac{\lambda}{2}\|\theta\|_2^2,
\end{equation}
and the parameter $\theta \in \Theta$ is represented as an $N$-dimensional vector $\theta = (\theta_1,\ldots,\theta_N)$. 

The function $l(\theta;z_i)$ represents a loss associated with each data point, 
such as a negative log-likelihood, and the term $\tfrac{\lambda}{2}\|\theta\|_2^2$  corresponds to the weight decay. The sum $\gL$ corresponds to a Hamiltonian in statistical mechanics. 
More generally, regularization can be incorporated into the measure over $\theta$, which would be more suitable when the regularization is more complicated, or the parameter is discrete. The parameter $\beta > 0$ is an inverse temperature. 
The case $\beta=1$ corresponds to the generalized Bayesian inference based on the score $l(\theta;z_i)$, which reduces to ordinary Bayesian inference when $l$ is the negative log-likelihood.
On the other hand, in the limit $\beta \to \infty$ the distribution concentrates on the minimizers of 
$\gL(\theta;\Dn)$, which corresponds to empirical risk minimization.

From this distribution we can consider a random variable $\hat{\theta}(\Dn)$ sampled as
\begin{equation}
  \hat{\theta}(\Dn) \sim p^\beta(\cdot \mid \Dn).
  \label{eq: sample from boltzmann distribution}
\end{equation}
The aim of statistical mechanics analysis is to study how $\hat{\theta}(\Dn)$ fluctuates 
according to thermal noise and quenched randomness, and how the statistical averages with respect to this distribution behave. 
In this way one can predict the behavior of quantities of interest in inference and learning, such as the prediction error.

We are particularly interested in the training error $\epsilon_{\rm tr}$, the prediction error $\epsilon_{\rm pred}$, 
and their difference, the generalization gap $\delta\epsilon$, which are defined through an error function 
$\epsilon(\cdot; \cdot):\Theta\times\gZ\to\mathbb{R}$ as
\begin{align}
  \epsilon_{\rm tr}(\Dn) &= \frac{1}{n}\sum_{i=1}^n \langle \epsilon(\theta;z_i) \rangle^{\beta ,n} \\
  \epsilon_{\rm pred}(\Dn) &= \mathbb{E}_{z\sim \pdata}\left[
    \langle \epsilon(\theta;z) \rangle^{\beta,n}
  \right] \\
  \delta\epsilon(\Dn) &= \epsilon_{\rm pred}(\Dn) - \epsilon_{\rm tr}(\Dn).
\end{align}

Here $\langle \;\cdot\; \rangle^{\beta,n}$ denotes the average with respect to 
the Boltzmann distribution \eqrefflat{eq: boltzmann}. 
These quantities depend on the training dataset $\Dn$ and are therefore random. 
It is often of interest to study their average behavior, such as
\begin{align}
  \bar{\epsilon}_{\rm tr} &= \mathbb{E}_{\Dn}\left[ \epsilon_{\rm tr}(\Dn)\right]
  \label{eq: average training error}
  \\
  \bar{\epsilon}_{\rm pred} &= \mathbb{E}_{\Dn}\left[ \epsilon_{\rm pred}(\Dn)\right]
  \label{eq: average prediction error}
  \\
  \bar{\delta\epsilon} &= \mathbb{E}_{\Dn}\left[ \delta\epsilon(\Dn)\right].
  \label{eq: average generalization gap}
\end{align}

We also remark that the error metric $\epsilon$ does not have to coincide with the loss function $l$ used in the definition of $\gL$.

\section{Replica Method}
\label{section: replica method}

To systematically investigate the fluctuations of $\hat{\theta}(\Dn)$ defined in \eqrefflat{eq: sample from boltzmann distribution} with respect to thermal noise and quenched randomness, it is useful to consider the replicated system defined for natural numbers $r=1,2,\ldots$ as a density on $\Theta^r$:
\begin{equation}
  p_{r}^{\beta,n}(\btheta) 
  = \frac{1}{\Xi_r^{\beta,n}} \, \E_{\Dn}\left[
    \prod_{a=1}^r e^{-\beta \gL(\theta^a;\Dn)}
  \right], \quad  \theta^a\in\Theta, a\in[r],
  \label{eq: replicated system canonical}
\end{equation}
where $\Xi_r^{\beta,n}$ is normalization constant and $[r]\equiv\{1,2,\ldots,r\}$. We also use the shorthand notation $\btheta \in \Theta^r$ to denote the concatenated vector of replicas, 
that is, the long vector obtained by stacking 
$\theta^1,\ldots,\theta^r$ vertically:
\begin{equation}
  \btheta = \begin{pmatrix}
    \theta^1 \\ \theta^2 \\ \vdots \\ \theta^r
  \end{pmatrix}.
\end{equation}
 Since the definition already involves an average over the dataset $\Dn$, this distribution is no longer conditioned on $\Dn$. 
However, the fluctuation of $\hat{\theta}(\Dn)$ with respect to quenched randomness is encoded in the correlations among replicas, which can be accessed through the correlation functions such as $\E_{r}^{\beta,n}[\theta_i^1 \theta_i^2]$, where $\E_{r}^{\beta,n}[ \cdot ]$ denotes the average with respect 
to the replicated system \eqrefflat{eq: replicated system canonical}. By extrapolating these quantities to real values of $r$, one can obtain the desired information. 

\subsection{Grand canonical formalism}
\label{subsection: grand canonical}
For subsequent analysis, it is convenient to rewrite the replicated system \eqrefflat{eq: replicated system canonical} as follows. The expectation with respect to $\Dn$ is taken over i.i.d.\ sampling $z_i \sim \pdata, \, i \in [n]$. In the i.i.d.\ case, this averaging is equivalent to sampling $n$ data points with replacement from a sufficiently large dataset $\tilde{D}=\{\tilde{z}_i\}_{i=1}^{\tilde{n}}, \, \tilde{z}_i \sim \pdata$, since the empirical distribution of a sufficiently large dataset converges to the true distribution $\pdata$ and can approximate it with arbitrary accuracy. This is in the same spirit as the bootstrap method developed by Efron \cite{10.1214/aos/1176344552, efron1994introduction}. Let $\tilde{c}_i \in [\tilde{n}]\cup\{0\}$ denote the number of times
$\tilde{z}_i$ is sampled. Then
\begin{equation}
  \E_{\Dn}\left[\prod_{a=1}^r e^{-\beta \gL(\theta^a;\Dn)}\right]
  \simeq  \E_{\tilde{\bm{c}}}\left[
    \prod_{i=1}^{\tilde{n}} e^{-\tilde{c}_i \beta \sum_{a=1}^r l(\theta^a;\tilde{z}_i)}
  \right] e^{-\tfrac{\beta\lambda}{2}\sum_{a=1}^r \|\theta^a\|_2^2},
\end{equation}
with $\tilde{\bm{c}}=(\tilde{c}_1,\ldots,\tilde{c}_{\tilde{n}})$.
In general, $\tilde{\bm{c}}$ follows a multinomial distribution, but for
$n,\tilde{n}\gg 1$ it can be approximated by independent Poisson
variables, $\tilde{c}_i \sim_{\rm i.i.d.} {\rm Poisson}(n/\tilde{n})$.
By taking the expectation with respect to the Poisson variables $\tilde{c}_i$ and taking the limit $\tilde{n}\to\infty$, the replicated system \eqrefflat{eq: replicated system canonical} can be rewritten in the following grand canonical (GC) form:
\begin{align}
  p_{r,\gc}^{\beta, n}(\bm{\theta})
  &= \frac{1}{\Xi_{r,\gc}^{\beta,n}}
    e^{n H(\btheta)},
  \label{eq: replicated system grand canonical} \\
  H(\btheta)
  &= \E_{z\sim\pdata}\left[e^{-\beta \sum_{a=1}^r l(\theta^a;z)}\right] -\tfrac{\beta\lambda}{2n}\sum_{a=1}^r \|\theta^a\|_2^2.
\end{align}

In fact, this rewriting is similar to the transform from canonical ensemble to the grand canonical ensemble in statistical mechanics where the number of data points, instead of the number of particles, fluctuates.
Hence, we refer to \eqrefflat{eq: replicated system grand canonical} as the grand canonical replicated system.
See Appendix~\ref{appendix: grand canonical} for a derivation more directly emphasizing the formal analogy with the grand canonical ensemble.
When computing averages, either \eqrefflat{eq: replicated system canonical} or \eqrefflat{eq: replicated system grand canonical} can be used in principle, but as we shall see, the GC formalism often simplifies the analysis.

\subsection{General formulae for errors}
Before proceeding to the variational approximation of the replicated system, we here summarize general formulas for error metrics \eqrefflat{eq: average training error}--\eqrefflat{eq: average generalization gap} 
based on the replica method. 
For this, we rewrite the normalization factor of the Boltzmann distribution, which appears in the definitions of the error metrics, as 
$1/Z^{\beta,n}(\Dn)=\lim_{r\to0}(Z^{\beta,n}(\Dn))^{r-1}$, 
and note that the normalization constant of the replicated system converges to one as $r\to0$. 
As is common in replica calculations, after symmetrization over replica indices the error metrics can be expressed as
\begin{align}
  \bar{\epsilon}_{\rm tr} &= \lim_{r\to0}\lim_{\gamma\to0}\frac{d}{d\gamma}\log \E_{\Dn}\left[
    \int e^{-\beta \sum_{a=1}^r\gL(\theta^a;\Dn) +\frac{\gamma}{nr}\sum_{i=1}^n\sum_{a=1}^r \epsilon(\theta^a;z_i)}d\theta^1\ldots d\theta^r
  \right],
  \\
  \bar{\epsilon}_{\rm pred} &= \lim_{r\to0}\E_{z\sim\pdata}\left[
    \E_{r}^{\beta,n}\left[
      \frac{1}{r}\sum_{a=1}^r \epsilon(\theta^a;z)
    \right]
  \right],
\end{align}
where $\lim_{r\to0}$ should be interpreted in the sense of the replica trick, as the extrapolation of results from integer $r$ to zero. Recall that $\E_{r}^{\beta,n}[\cdot]$ is the expectation with respect to the replicated system \eqrefflat{eq: replicated system canonical}. 

By rewriting the dataset average as in the derivation of the GC formalism, and replacing the expectation $\E_{r}^{\beta,n}$ with $\E_{r,\gc}^{\beta,n}$, we obtain the following compact forms:
\begin{align}
  \bar{\epsilon}_{\rm tr} &= \E_{z\sim\pdata}\left[
    \lim_{r\to0}\frac{1}{r}\sum_{a=1}^r \E_{r, \gc}^{\beta,n}\left[\epsilon(\theta^a;z)e^{-\beta\sum_{b=1}^r l(\theta^b;z)}\right]
  \right],
  \\
  \bar{\epsilon}_{\rm pred} &= \E_{z\sim\pdata}\left[
    \lim_{r\to0}\frac{1}{r}\sum_{a=1}^r \E_{r, \gc}^{\beta,n}\left[\epsilon(\theta^a;z)\right]
  \right],
\end{align}
which further yield
\begin{equation}
  \bar{\delta\epsilon} = \E_{z\sim\pdata}\left[
    \lim_{r\to0}\frac{1}{r}\sum_{a=1}^r \E_{r, \gc}^{\beta,n}\left[
      (1-e^{-\beta\sum_{b=1}^r l(\theta^b;z)})\epsilon(\theta^a;z)
    \right]
  \right].
  \label{eq: genralization gap replica}
\end{equation}

In these expressions, the expectation with respect to the GC replicated system plays the role of a cavity bias at the data point $z$.  In particular, the formula for the training error shows that the factor $e^{-\beta\sum_{a=1}^r l(\theta^a;z)}$ appears as a bias due to the inclusion of the evaluation point in the training dataset.

\subsubsection{Remark (link to PCIC/WAIC)}
Here we briefly note a connection with information criteria, which are estimators for the generalization gap \cite{konishi2008information}. If we expand the factor $e^{-\beta\sum_{a=1}^r l(\theta^a;z)}$ in a Taylor series 
and analytically continue as $r\to 0$ without considering replica symmetry breaking fields, we obtain
\begin{equation}
  \bar{\delta\epsilon} 
  = \E_{z\sim\pdata}\Bigl[\,
    \beta \left(
        \langle l(\theta;z)\epsilon(\theta;z) \rangle^{\beta,n} 
        - \langle l(\theta;z)\rangle^{\beta,n}\langle \epsilon(\theta;z)\rangle^{\beta,n} 
      \right)
  \Bigr] + \ldots.
\end{equation}
If the expectation over $z$ is replaced by the empirical average over the training data, the first-order term coincides with the posterior covariance information criterion (PCIC) 
\cite{10.1162/neco_a_01592}, and for $l=\epsilon$ it coincides with widely applicable information criterion (WAIC) \cite{watanabe2013widely}. In particular, when $l=\epsilon$, the expansion corresponds to a cumulant expansion with respect to $\beta l$.

Unfortunately, it is difficult to obtain useful bounds on the higher-order terms from this representation, because simple bounds such as $1-e^{-x}\le x$, valid for $r\in\mathbb{N}$, are not guaranteed to yield meaningful bounds after analytic continuation to $r\to0$. Therefore, while this correspondence is interesting, its practical usefulness remains unclear to the author at present.

\section{Variational Gaussian Approximation (VGA)}
\label{section: variational gaussian}
The replicated systems \eqrefflat{eq: replicated system canonical} and \eqrefflat{eq: replicated system grand canonical} are in general intractable unless the data generating distribution $\pdata$ is simple enough. 
To proceed, we adopt a variational approximation in GC formalism. We introduce a trial Hamiltonian $\tilde{H}$, which is determined by the stationarity conditions of the variational free energy
\begin{equation}
  \gF_{r,\gc}^{\beta,n} = \log \tilde{\Xi}_{r,\gc}^{\beta,n} 
  + n \, \Ec\left[ H(\btheta) - \tilde{H}(\btheta) \right],
  \label{eq: variational free energy}
\end{equation}
where $\Ec[\cdot]$ denotes the average with respect to the approximate replicated system $e^{n \tilde{H}(\bm{\theta})}/\tilde{\Xi}_{r,\gc}^{\beta,n}$, where $\tilde{\Xi}_{r,\gc}^{\beta,n}$ is the normalization constant. 
The derivation of this variational free energy follows the standard perturbative variational principle \cite{Parisi:111935,opper2001advanced,pml2Book} and, for integer $r$, it provides a bound on the true free energy. 
However, in the limit $r\to 0$, its leading term $\lim_{r\to 0} r^{-1}\gF_{r,\gc}^{\beta,n}$ is not necessarily guaranteed to bound the true free energy $-\E_{\Dn}[\log Z^{\beta,n}(\Dn)]$. 
Hence, instead of naively maximizing or minimizing with respect to the parameters of the trial Hamiltonian, we have to consider stationary conditions. Appendix~\ref{appendix: simplest case} illustrates a simple example in which the variational parameters are at a saddle point rather than a maximum or minimum.
It should be noted that the stationarity conditions are considered solely from the requirement of choosing the trial Hamiltonian optimally. Hence, unlike in many derivations of exact solutions, taking the thermodynamic limit is not necessary here, although one could consider an appropriate limit later.

The simplest trial Hamiltonian that incorporates interactions between parameters is a quadratic form:
\begin{equation}
  \tilde{H}(\btheta) 
  = -\frac{1}{2}(\bm{\theta}-\bm{m})^\top Q^{-1}(\bm{\theta}-\bm{m}),
  \label{eq: trial hamiltonian}
\end{equation}
where $\bm{m}=[m_i^a]_{i\in[N],\,a\in[r]}\in\R^{Nr}$ is the concatenated vector  formed by stacking $m^a$, and 
$Q=[Q_{ij}^{ab}]_{i,j\,\in\,[N],\,a,b\,\in\,[r]}\in\R^{Nr\times Nr}$ is a covariance matrix. We remark that, unlike in the conventional spin-glass literature where $Q$ typically denotes second moments (the so-called overlaps), here $Q$ corresponds to the covariance matrix.
For now, we consider a general structure of the mean and covariance, though more specific assumptions may be adopted depending on prior knowledge or computational constraints.

With the trial Hamiltonian \eqrefflat{eq: trial hamiltonian}, the variational free energy takes the following form, up to an additive constant,
\begin{align}
  \gF_{r,\gc}^{\beta,n} &= \gF_{\rm ent} + \gF_{\rm reg} + \gF_{\rm int},
  \label{eq: variational free energy gaussian} \\
  \gF_{\rm ent} &= \frac{1}{2}\log \det (Q/n), \\
  \gF_{\rm reg} &= -\frac{\beta \lambda}{2n}\sum_{a=1}^r\sum_{i=1}^N 
    \left( (m_i^a)^2 + Q_{ii}^{aa} \right), \\
  \gF_{\rm int} &= n \, \E_{z\sim\pdata}\left[
    \Ec\left[ e^{-\beta \sum_{a=1}^r l(\theta^a;z)} \right]
  \right].
\end{align}

In the following, the variational parameters $\bm{m}$ and $Q$ are determined under the replica-symmetric (RS) assumption by imposing the stationarity conditions of \eqrefflat{eq: variational free energy gaussian}. 
We then use the approximate distribution $e^{n\tilde{H}(\bm{\theta})}/\tilde{\Xi}_{r,\gc}^{\beta,n}$ instead of the true replicated system \eqrefflat{eq: replicated system grand canonical} to compute relevant quantities. 
For example, the generalization gap can be evaluated as 
\begin{equation}
  \bar{\delta\epsilon} \simeq 
  \E_{z\sim \pdata}\left[
    \lim_{r\to 0} \Ec\left[
      \frac{1}{r}\sum_{a=1}^r \epsilon(\theta^a;z)\,
      \bigl(1 - e^{-\beta \sum_{b=1}^r l(\theta^b;z)}\bigr)
    \right]
  \right].
\end{equation}

It is worth emphasizing that the use of the GC formulation yields the interaction term $\gF_{\rm int}$ in a form where the expectation over the data distribution $z\sim\pdata$ is taken \emph{after} averaging the factor $e^{-\beta \sum_{a=1}^r l(\theta^a;z)}$ with respect to the approximate distribution. 
If we instead use the original formulation \eqrefflat{eq: replicated system canonical}, one would have to deal with the term $\log \E_{z\sim\pdata}[e^{-\beta \sum_{a=1}^r l(\theta^a;z)}]$ 
inside the average with respect to the approximate distribution. 
Such a factor is difficult to handle unless the expectation over $\pdata$ can be computed analytically.
This makes progress almost impossible unless the average over $\pdata$ is analytically tractable. 
By contrast, in the present formulation one may obtain a tractable expression for the interaction term by averaging with respect to the approximate distribution, provided that the trial Hamiltonian is sufficiently simple.

In practice, we derive the stationarity conditions while keeping the expectation over $\pdata$ explicit, and when needed we approximate it with the empirical average of the observed data.
This makes it possible to obtain approximate formulae that apply to general datasets.

\section{General Results}
\label{section: general results}

In this section, we summarize general consequences of the RS parameterization that are largely independent of the specific form of the loss function $l$.

We consider the RS parameterization
\begin{align}
  Q_{ij}^{ab} &= q_{ij} + \delta_{ab}\frac{\chi_{ij}}{\beta}, \\
  m_i^a &= m_i.
\end{align}

This is equivalent to assuming that the estimator $\hat{\theta}(\Dn)$ behaves as an effective random variable of the form
\begin{equation}
  \begin{split}
      &\hat{\theta}(\Dn) = m + \xi + \eta, 
      \\
      &\xi \sim \gN(0,q/n), \quad 
      \eta \sim \gN(0,\chi/(\beta n)),
  \end{split}
\end{equation}
where $\xi$ represents quenched fluctuations and $\eta$ represents thermal fluctuations. Also, $m$ is a parameter that explains how the training data breaks the symmetry of the model parameter.

Conditioned on $\xi \sim \gN(0,q/n)$ and $z\sim\pdata$, we introduce the auxiliary densities on $\Theta$:
\begin{align}
  \phi(\theta \mid \xi, z) 
  &= \frac{1}{Z_\phi}
  e^{
    -\frac{\beta n}{2}(\theta - m - \xi)^\top \chi^{-1} (\theta - m - \xi)
    - \beta l(\theta;z)
  },
  \label{eq: cavity + measurement general}
  \\
  \phiz(\theta \mid \xi) 
  &= \frac{1}{Z_{\phiz}}
  e^{
    -\frac{\beta n}{2}(\theta - m - \xi)^\top \chi^{-1} (\theta - m - \xi)
  }.
  \label{eq: cavity general}
\end{align}
We denote expectations with respect to $\phi$ and $\phiz$ by 
$\langle \cdot \rangle_{\phi}$ and $\langle \cdot \rangle_{\phiz}$, respectively.

For any function $g:\Theta\times\gZ \to \R$, apart from the difference between the factors $n-1$ and $n$, we obtain
\begin{equation}
  \E_{\Dn}\left[g(\langle \theta \rangle^{\beta,n}, z_i)\right] 
  \simeq \E_{z\sim\pdata,\,\xi\sim \gN(0,q/n)}\left[
    g(\langle \theta \rangle_\phi, z)
  \right],
  \label{eq: in sample prediction general}
\end{equation}
for $z_i\in\Dn$, while for a fresh sample $\tilde{z}\sim\pdata$ independent of $\Dn$,
\begin{equation}
  \E_{\Dn,\tilde{z}}\left[g(\langle \theta \rangle^{\beta,n}, \tilde{z})\right] 
  \simeq \E_{z\sim\pdata,\,\xi\sim \gN(0,q/n)}\left[
    g(\langle \theta \rangle_{\phiz}, z)
  \right].
  \label{eq: cavity prediction general}
\end{equation}
These suggest that $\phiz$ represents the thermal fluctuations of the cavity bias in the absence of a specific data point, while including $z$ corresponds to tilting by the factor $e^{-\beta l(\theta;z)}$.

In the zero-temperature limit $\beta \to \infty$, these reduce to
\begin{align}
  \E_{\Dn}\left[g(\langle \theta \rangle^{\beta,n}, z_i)\right] 
  &\to \E_{z,\xi}\left[ g(\theta^\ast, z) \right], \\
  \E_{\Dn,\tilde{z}}\left[g(\langle \theta \rangle^{\beta,n}, \tilde{z})\right] 
  &\to \E_{z,\xi}\left[ g(m+\xi, z) \right],
\end{align}
where
\begin{equation}
  \theta^\ast = \arg\min_{\theta}
    \frac{n}{2}(\theta - m - \xi)^\top \chi^{-1} (\theta - m - \xi) 
    + l(\theta;z).
    \label{eq: theta ast general}
\end{equation}

Consequently, the averaged generalization gap takes the compact form
\begin{align}
  \bar{\delta \epsilon}
  &= \E_{z,\xi}\left[
    \langle \epsilon(\theta;z) \rangle_{\phiz}
    - \langle \epsilon(\theta;z) \rangle_{\phi}
  \right] \\
  &\overset{\beta\to\infty}{\longrightarrow}
  \E_{z,\xi}\left[
    \epsilon(m+\xi;z) - \epsilon(\theta^\ast;z)
  \right].
\end{align}

\subsection{Stationarity conditions}
\label{subsection: stationarity condition general}
The variational parameters $m,q,\chi$ are determined by the following stationarity conditions:
\begin{align}
  0 &= \lambda m + n \E_{z\sim\pdata}\left[
    \langle \nabla_{\theta} l(\theta;z) \rangle_{\phi}
  \right], 
  \label{eq: stationarity m general}\\
  \chi^{-1} q \chi^{-1} &= \E_{z\sim\pdata}\left[
    \langle \nabla_{\theta} l(\theta;z) \rangle_{\phi}
    \langle \nabla_{\theta} l(\theta;z) \rangle_{\phi}^\top
  \right], \\
  \chi^{-1} &= \frac{\lambda}{n} I_N + \E_{z\sim\pdata}\left[
    \left.
      \nabla_{\gamma}\langle \nabla_{\theta} l(\theta+\gamma;z) \rangle_{\phi_\gamma}^\top
    \right|_{\gamma=0}
  \right],
\end{align}
where
\begin{equation}
  \phi_{\gamma}(\theta \mid \xi, z) \propto 
  e^{
    -\frac{\beta n}{2}(\theta-m-\xi)^\top \chi^{-1} (\theta-m-\xi)
    -\beta l(\theta+\gamma;z)
  }.
\end{equation}
These conditions correspond, respectively, to the stationarity of the population loss gradient, a (non-centered) covariance-like quantity for the gradients, and a regularized Hessian-like quantity. They are, however, evaluated under thermal averages, and therefore do not coincide with naive population quantities when $n$ is finite.

It may be useful to comment on the stationarity condition for $m$.   Suppose that in the condition \eqrefflat{eq: stationarity m general} both $q$ and $\chi$ vanish.  In this case, the stationary condition reduces to  
$
0 = \lambda m + n\,\E_{z}[\nabla_m l(m;z)]
$,
which corresponds to the stationary point of the population loss, i.e., an ideal estimator. However, in general, this is of course not the case. Hence, one important question is how this ideal condition is modified by the thermal fluctuations $\chi$ and the quenched randomness $q$. To reveal the structure of the solution, one typically needs either simplifying assumptions or numerical analysis, yet examining the form of the governing equations already provides insight into how well the estimator learns the average direction encoded by $m$. We will return to this point in subsubsection~\ref{subsubsection: double descent}.

\subsection{Asymptotic properties}

We now examine the asymptotic properties at $n\to\infty$. To simplify the discussion, we consider $\beta \to \infty$ and assume that the Hessian of the population loss at the stationary point is positive definite. Considering the perturbative expansion of $\theta^\ast$ in $1/n$ as 
\begin{equation}
  \theta^\ast = \theta_0^\ast + \frac{1}{n}\theta_1^\ast + \cdots,
\end{equation}
and straightforwardly expanding the stationary condition for the optimization problem \eqrefflat{eq: theta ast general}
\(
n \chi^{-1}(\theta^\ast - m - \xi) + \nabla_\theta l(\theta^\ast;z) = 0,
\)
we obtain
\begin{equation}
  \theta_0^\ast = 0, 
  \qquad
  \theta_1^\ast = \chi_0^{-1}\nabla_\theta l(m_0;z),
\end{equation}
where $m_0$ satisfies the stationary condition of the population loss
\begin{equation}
  0 = \E_{z\sim\pdata}\left[\nabla_\theta l(m_0;z)\right],
\end{equation}
and $\chi_0$ denotes the inverse Hessian of the population loss at $m_0$:
\begin{equation}
  \chi_0 = 
  \left(
    \E_{z\sim\pdata}[
      \nabla_\theta l(m_0;z)\nabla_\theta l(m_0;z)^\top
    ]
  \right)^{-1}.
\end{equation}
In this regime, the generalization gap is given by
\begin{align}
  \bar{\delta \epsilon}
  = \frac{1}{n}\Tr\left[
    \chi_0 \, \E_{z\sim\pdata}[
      \nabla_{\theta}\epsilon(m_0;z) \nabla_{\theta} l(m_0;z)^\top
    ]
  \right] + \gO(n^{-2}),
\end{align}
which is precisely the Takeuchi Information Criterion 
\cite{takeuchi1976distribution,PhysRevA.45.6056,amari1993statistical,amari1995statistical}.
If the regularization parameter scales as $\lambda n$, then the condition for $m_0$ and $\chi_0^{-1}$ should be replaced by
\begin{align}
  0 &= \lambda m_0 + \E_{z\sim\pdata}[\langle
    \nabla_{\theta} l(m_0;z)
  \rangle],
  \\
  \chi_0^{-1} &= \lambda I_N + \E_{z\sim\pdata}[
    \nabla_\theta l(m_0;z)\nabla_\theta l(m_0;z)^\top
  ].
\end{align}

\section{Application to Linear Models}
\label{section: linear models}
As a concrete example, we consider linear regression. 
In this setting, the data consist of input–output pairs $(x_i, y_i)$ with $x_i \in \R^d$ and $y_i \in \R$, and the parameter vector corresponds to the regression coefficients.
For clarity, we denote the parameter by $w \in \R^d$ instead of the generic notation $\theta$.  
In linear regression, the input dimension coincides with the parameter dimension, so that $N=d$.  
For notational convenience, we explicitly indicate that the variational parameters are associated with $w$, writing them as $m_w$ and $Q_w$.  
With a slight abuse of notation, we replace the general loss $l(\theta;z)$ by a function $l:\R \times \R \to \R$ that depends on the response $y$ and the scaled inner product
\(
  \langle x, w \rangle \equiv d^{-1/2}\sum_{i=1}^d x_i w_i,
\)
so that in the linear regression setting the loss takes the form 
$l(y, \langle x, w\rangle)$.
To focus on the usual ridge-regularized empirical risk minimization, it is sufficient to consider the limit $\beta\to\infty$.

\subsection{General Properties}
We now consider the general properties of linear models.  
Under the RS parametrization
\(
  m_w^a = m_w, 
  \quad 
  Q_w^{ab} = q_w + \delta_{ab}\chi_w/\beta,
\)
the output of the model $\hat{f}(x) = \langle x, \hat{w}(\Dn)\rangle$ with $\hat{w}(\Dn) \sim p^\beta(\cdot \mid \Dn)$ can be effectively represented as the random variable
\begin{equation}
  \hat{f}(x) = m_f(x) + \xi_f + \eta_f,
\end{equation}
where
\begin{align}
  m_f(x) &= \langle x, m_w\rangle, \\
  \xi_f &\sim \gN(0, q_f(x)), \\
  \eta_f &\sim \gN(0, \chi_f(x)/\beta).
\end{align}
Here, $\xi_f$ accounts for quenched randomness and $\eta_f$ for thermal fluctuations, with the variances given by
\begin{align}
  q_f(x) &= \frac{1}{dn}\sum_{i,j=1}^d x_i x_j q_{w,ij}, \\
  \chi_f(x) &= \frac{1}{dn}\sum_{i,j=1}^d x_i x_j \chi_{w,ij}.
\end{align}
Thus, the model output is characterized by a Gaussian process whose mean and variance both depend on the input $x$.

Similar to \eqrefflat{eq: cavity + measurement general}–\eqrefflat{eq: cavity prediction general}, 
we can also describe the effective fluctuations of observables at training samples $z_i \in \Dn$ 
and at fresh samples $\tilde{z} \sim \pdata$ not contained in $\Dn$.  
Conditioned on $\xi_f$ and $z=(x,y)$, we introduce the distributions
\begin{align}
  \phi^{\rm lin}(f \mid \xi_f, z) 
  &= \frac{1}{Z_\phi^{\rm lin}}
  e^{
    -\frac{\beta (f - m_f(x)-\xi_f)^2}{2\chi_f(x)}
    - \beta l(y,f)
  }, \\
  \phi^{\rm lin}_{\backslash z}(f \mid \xi_f, x) 
  &= \frac{1}{Z_{\phi_{\backslash z}}^{\rm lin}}
  e^{
    -\frac{\beta (f - m_f(x)-\xi_f)^2}{2\chi_f(x)}
  }.
\end{align}
Here, $\phi^{\rm lin}$ and $\phi^{\rm lin}_{\backslash z}$ are effective descriptions of the thermal fluctuations 
of the output $\langle x, \hat{w}(\Dn)\rangle^{\beta,n}$ at $z \in \Dn$ and $z \notin \Dn$, respectively.  
The factor proportional to $e^{-\beta \frac{(f - m_f(x) - \xi_f)^2}{2\chi_f(x)}}$ 
captures the cavity bias due to all other data points except for $z$.

Using these distributions, for any function $g:\R\times\gZ \to \R$, we then obtain
\begin{align}
  \E_{\Dn}\left[g\left(\bigl\langle \langle x_i, w\rangle \bigr\rangle^{\beta,n}, z_i\right)\right] 
  &= \E_{z}\left[
    \E_{\xi_f\sim\gN(0,q_f(x))}\left[
      g\bigl(\langle f\rangle_{\phi^{\rm lin}}, z\bigr)
    \right]
  \right] \\
  &\overset{\beta\to\infty}{\longrightarrow} 
  \E_{z}\left[
    \E_{\xi_f\sim\gN(0,q_f(x))}\left[
      g(f^\ast, z)
    \right]
  \right], \\
  \E_{\Dn, \tilde{z}}\left[g\left(\bigl\langle \langle \tilde{x}, w\rangle \bigr\rangle^{\beta,n}, \tilde{z}\right)\right] 
  &= \E_{z}\left[
    \E_{\xi_f\sim\gN(0,q_f(x))}\left[
      g\bigl(\langle f\rangle_{\phi^{\rm lin}_{\backslash z}}, z\bigr)
    \right]
  \right] \\
  &\overset{\beta\to\infty}{\longrightarrow} 
  \E_{z}\left[
    \E_{\xi_f\sim\gN(0,q_f(x))}\left[
      g\bigl(m_f(x)+\xi_f, z\bigr)
    \right]
  \right],
\end{align}
where
\begin{equation}
  f^\ast = \arg\min_{f} 
  \frac{(f - m_f(x) - \xi_f)^2}{2\chi_f(x)} + l(y,f).
\end{equation}

Accordingly, if the error is measured by a function $\epsilon(\cdot,\cdot): \R \times \R \to \R$, the data-averaged generalization gap takes the form
\begin{align}
  \bar{\delta\epsilon} 
  &= \E_{z\sim\pdata}\left[
    \E_{\xi_f\sim\gN(0,q_f(x))}\left[
      \langle \epsilon(y,f)\rangle_{\phi^{\rm lin}_{\backslash z}}
      - \langle \epsilon(y,f)\rangle_{\phi^{\rm lin}}
    \right]
  \right] \\
  &\to \E_{z\sim\pdata}\left[
    \E_{\xi_f\sim\gN(0,q_f(x))}\left[
      \epsilon(y, m_f(x)+\xi_f)
      - \epsilon(y, f^\ast)
    \right]
  \right].
\end{align}

In the limit $\beta \to \infty$, the stationarity conditions take the following form, where $l^\prime$ denotes the derivative of $l$ with respect to its second argument:
\begin{align}
  0 &= \lambda m_w 
  + n\,\E_{z,\xi_f}\left[
    \frac{x}{\sqrt{d}}\, l^\prime \bigl(y, f^\ast(x)\bigr)
  \right], \\
  \chi_w^{-1} q_w \chi_w^{-1} 
  &= \E_{z,\xi_f}\left[
    \frac{xx^\top}{d}\,\bigl(l^\prime(y,f^\ast(x))\bigr)^2
  \right], \\
  \chi_w^{-1} 
  &= \frac{\lambda}{n}I_d 
  + \E_{z,\xi_f}\left[
    \frac{xx^\top}{d}\, 
    \frac{d}{d\gamma} l^\prime\bigl(y, f^\ast_\gamma(x) + \gamma\bigr)
  \right],
\end{align}
with
\begin{equation}
  f^\ast_\gamma = \arg\min_{f} 
  \frac{(f - m_f(x) - \xi_f)^2}{2\chi_f(x)} + l(y, f+\gamma).
\end{equation}

In practice, even when the data-generating distribution $\pdata$ is unknown, the expectations over $z$ that appear in the above expressions can be approximated by empirical averages over the obtained data. This allows the generalization gap to be estimated for arbitrary distributions.

\subsection{Linear regression}
\label{subsection: linear regression}

We now specialize to the case of squared loss, 
$l(y,f) = \tfrac{1}{2}(y-f)^2$, with the error metric 
$\epsilon(y,f) = (y-f)^2$.  
In this case, the stationarity conditions take the form
\begin{align}
  0 &= \lambda m_w + n\,\E_{z}\left[
    \frac{1}{1+\chi_f(x)}\left(
      -\frac{x}{\sqrt{d}}\,(y-m_f(x))
    \right)
  \right], 
  \label{eq: stationarity m}\\
  \chi_w^{-1}q_w\chi_w^{-1} &= \E_z\left[
    \frac{xx^\top}{d}\,\frac{1}{(1+\chi_f(x))^2}
      \left((y-m_f(x))^2 + q_f(x)\right)
  \right]
  \label{eq: stationarity q}, \\
  \chi_w^{-1} &= \frac{\lambda}{n}I_d + \E_z\left[
    \frac{xx^\top}{d}\,\frac{1}{1+\chi_f(x)}
  \right]
  \label{eq: stationarity chi}.
\end{align}

The corresponding error metrics are given by
\begin{align}
  \bar{\epsilon}_{\rm tr} &= \E_{z}\left[
    \frac{1}{(1+\chi_f(x))^2}\bigl(
      (y-m_f(x))^2 + q_f(x)
    \bigr)
  \right], \\
  \bar{\epsilon}_{\rm pred} &= \E_{z}\left[
    (y-m_f(x))^2 + q_f(x)
  \right], \\
  \bar{\delta\epsilon} &= \E_{z}\left[
    \Bigl(1-\frac{1}{(1+\chi_f(x))^2}\Bigr)\bigl(
      (y-m_f(x))^2 + q_f(x)
    \bigr)
  \right].
\end{align}

In the case of linear regression with squared loss, the equation determining $m_w$ is independent of $q_w$, but it is affected by the susceptibility $\chi_f(x) > 0$. Specifically, the effective gradient information is suppressed 
by the factor $1/(1+\chi_f(x)) < 1$.

\subsubsection{Remark on double descent}
\label{subsubsection: double descent}
As discussed in Subsection~\ref{subsection: stationarity condition general}, understanding how fluctuations affect the extraction of information is crucial for analyzing the estimator. 
A particularly instructive case is the weak regularization limit $\lambda \to +0$. 
In this regime, the estimator reduces to the $\ell_2$ norm interpolator, which interpolates the training data and selects the solution of minimal $\ell_2$ norm in the underdetermined case. 
Thanks to the simplicity of the stationarity conditions in linear regression, this setting allows us to see more transparently how fluctuations affect the extraction of signal.

To this end, we assume that inputs are normalized as $\|x\|_2^2 = d$ and that their components are i.i.d.  
It is then natural to adopt a simplified ansatz $q_w = \bar{q}_w I_d, \; \chi_w = \bar{\chi}_w I_d$, 
so that $\chi_f(x) = \bar{\chi}_w/n$ becomes independent of $x$ (extension to more general correlated features is straightforward). 
Writing $\alpha = n/d$, we obtain
\begin{equation}
  \frac{1}{1+\chi_f(x)} = 
  \begin{cases}
    1-\frac{1}{\alpha} + \gO(\lambda), & d < n, \\
    \frac{\lambda}{1-\alpha} + \gO(\lambda^2), & n < d.
  \end{cases}
\end{equation}
Then, the stationarity equation for $m_w$ becomes
\begin{align}
  0 &= 
  \E_{z\sim\pdata}\left[-\frac{x}{\sqrt{d}}(y-m_f(x))\right]
  + \gO(\lambda), 
  & d<n, \\
  0 &= m_w + \frac{n}{1-\alpha}
  \E_{z\sim\pdata}\left[-\frac{x}{\sqrt{d}}(y-m_f(x))\right]
  + \gO(\lambda), 
  & n<d.
\end{align}
When $d<n$, we have $\chi_f(x)=\gO(1)$, so the signal term remains of order one in the weakly regularized limit, and $m_w$ approaches the ideal estimator.  
In contrast, when $n<d$, $\chi_f(x)=\gO(1/\lambda)$ diverges, causing the signal term to vanish at order $\gO(\lambda)$ and thus introducing a bias even when $\lambda \to +0$.  
Although the precise form of the solution requires a more detailed analysis, the structure of the stationarity equations already reveals how such biases arise.

Similarly, the averaged generalization gap can be estimated as
\begin{equation}
  \bar{\delta\epsilon} = \overline{\rm RSS}\times
  \begin{cases}
    \frac{1}{\alpha} + \frac{1}{\alpha-1} + \gO(\lambda), & d<n, \\
    \frac{1}{1-\alpha} + \gO(\lambda), & n<d,
  \end{cases}
\end{equation}
where $\overline{\rm RSS} = \E_{z\sim\pdata}[(y-m_f(x))^2]$.  
Thus, as long as the residual is nonzero, the generalization gap diverges at $d=n$, causing the double descent phenomenon. 
While this result is well known in solvable teacher–student scenarios \cite{krogh_simple_nodate, hastie2022surprises}, the present variational framework reveals that the phenomenon can be understood without specifying a particular teacher model.

\subsection{Application to concrete problems}

The most notable feature of the present formulation is that the parameters of the trial Hamiltonian can be determined adaptively for a given data-generating distribution. 
By approximating the population average $\E_{z \sim \pdata}$ using some data, the method can be applied not only to synthetic data with analytically tractable expectations, but also to real datasets where the true distribution is unknown.  In this subsection, we illustrate this point with several concrete settings. 

For each case, we approximate the average $\E_z[\gO(z)]$ of a statistic $\gO(z)$ by the empirical average over the data of size $n_0$ as $n_0^{-1}\sum_{i=1}^{n_0}\gO(z_i)$. Using this approximation with fixed $n_0$, we numerically solve the stationarity conditions \eqrefflat{eq: stationarity m}-\eqrefflat{eq: stationarity chi} to determine $m_w, q_w, \chi_w$, and then predict the generalization error $\bar{\epsilon}_{\rm pred}$.  That is, we approximate $\E_z$ using a dataset of size $n_0$, and predict the learning behavior when trained on data of size $n$. It should be noted that, in general, $n_0$ is independent of $n$.  Given $n_0$, $n$ is merely a control parameter. Thus, both $n \le n_0$ and $n \ge n_0$ are possible. Of course, the accuracy of the final result depends on $n_0$, as discussed in subsubsection \ref{subsubsection: real-world data}.

We also remark that even with moderately large datasets, quantifying the fluctuations represented by $\chi_w$ and $q_w$ is not at all trivial without repeated experiments, yet the VGA provides a principled framework to estimate them.

\begin{figure*}[t]
  \captionsetup[subfloat]{captionskip=-1truemm}
    \centering
    \subfloat[linear, $\lambda=10^{-2}$]{
      \includegraphics[width=0.23\textwidth]{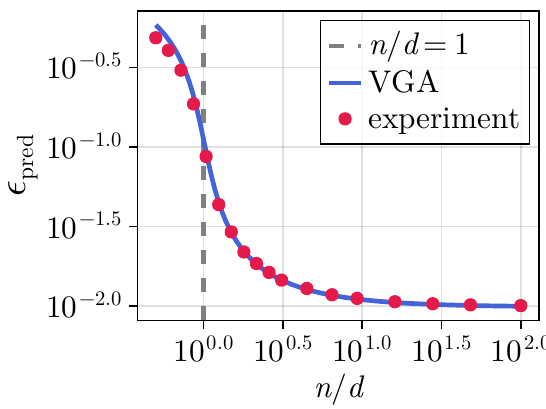}
    }\hfill
    \subfloat[linear, $\lambda=10^{-4}$]{
      \includegraphics[width=0.23\textwidth]{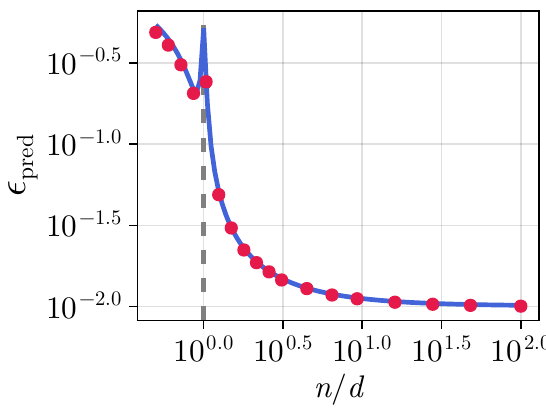}
    }\hfill
      \subfloat[linear, $\lambda=10^{-2}$]{
      \includegraphics[width=0.23\textwidth]{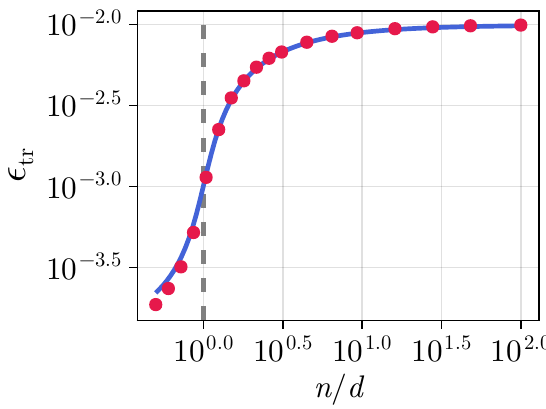}
    }\hfill
    \subfloat[linear, $\lambda=10^{-4}$]{
      \includegraphics[width=0.23\textwidth]{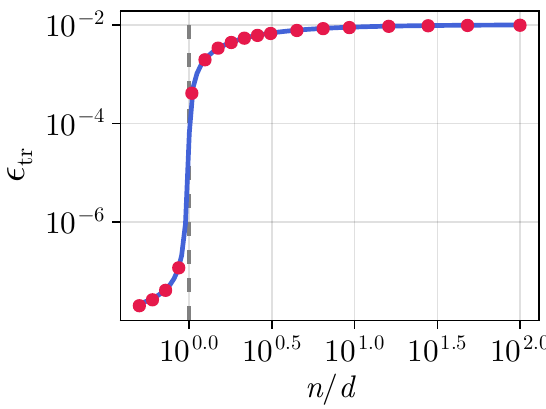}
    }\\
    \subfloat[2NN (relu), $\lambda=10^{-2}$]{
      \includegraphics[width=0.23\textwidth]{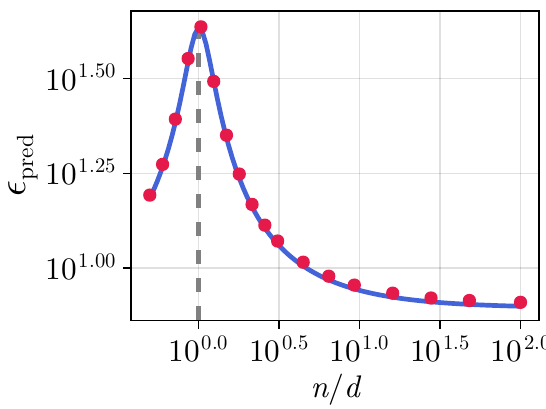}
    }\hfill
    \subfloat[2NN (relu), $\lambda=10^{-4}$]{
      \includegraphics[width=0.23\textwidth]{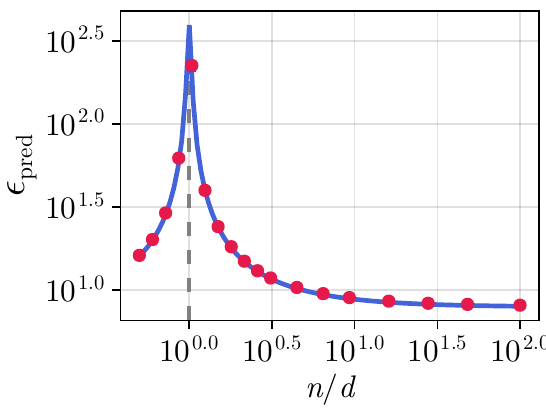}
    }\hfill
    \subfloat[2NN (relu), $\lambda=10^{-2}$]{
      \includegraphics[width=0.23\textwidth]{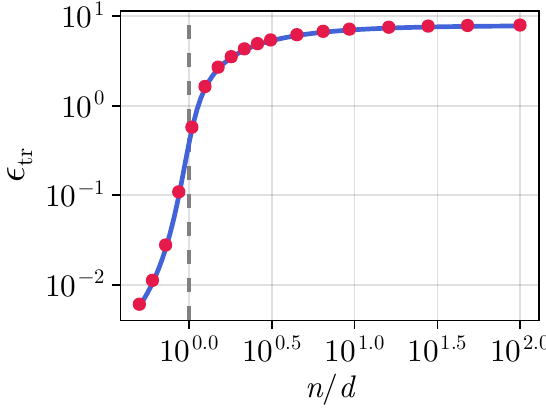}
    }\hfill
    \subfloat[2NN (relu), $\lambda=10^{-4}$]{
      \includegraphics[width=0.23\textwidth]{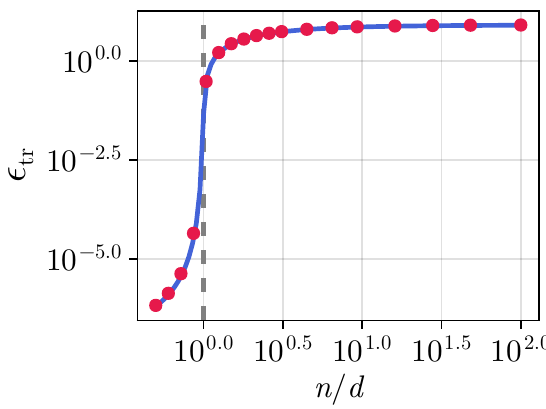}
    }\\
    \subfloat[2NN (tanh), $\lambda=10^{-2}$]{
      \includegraphics[width=0.23\textwidth]{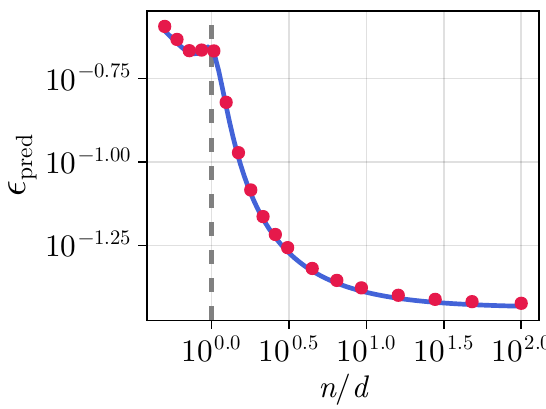}
    }\hfill
    \subfloat[2NN (tanh), $\lambda=10^{-4}$]{
      \includegraphics[width=0.23\textwidth]{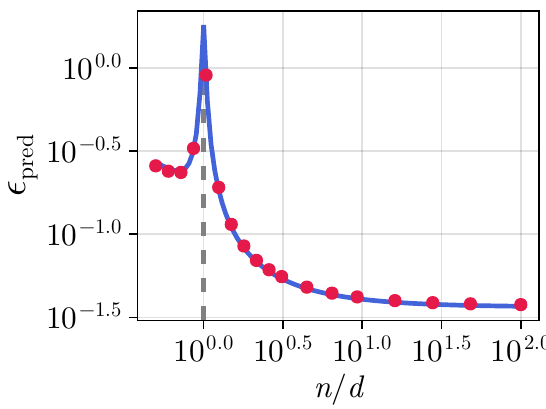}
    }\hfill
      \subfloat[2NN (tanh), $\lambda=10^{-4}$]{
      \includegraphics[width=0.23\textwidth]{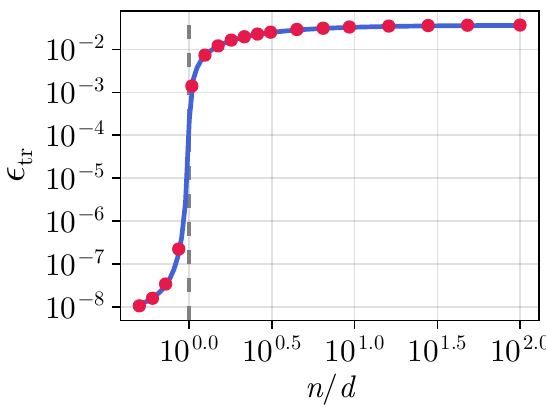}
    }\hfill
    \subfloat[2NN (tanh), $\lambda=10^{-2}$]{
      \includegraphics[width=0.23\textwidth]{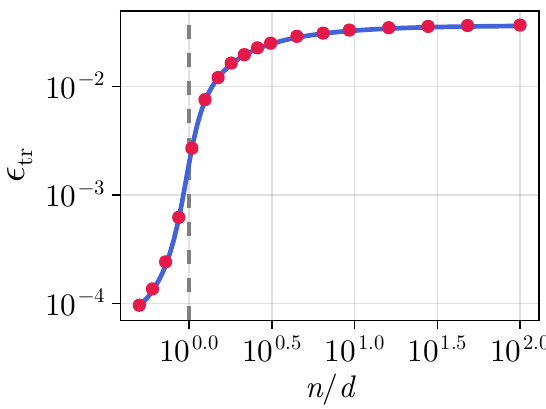}
    }\\
    \subfloat[YP MSD, $\lambda=10^{-2}$]{
      \includegraphics[width=0.23\textwidth]{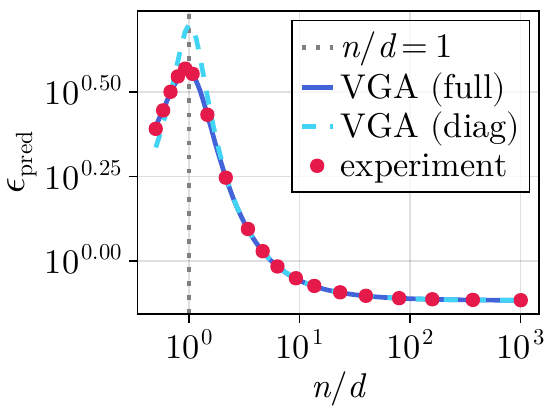}
    }\hfill
    \subfloat[YP MSD, $\lambda=10^{-4}$]{
      \includegraphics[width=0.23\textwidth]{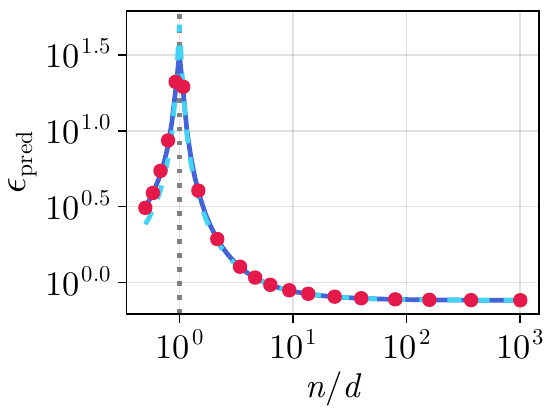}
    }\hfill
    \subfloat[YP MSD, $\lambda=10^{-2}$]{
      \includegraphics[width=0.23\textwidth]{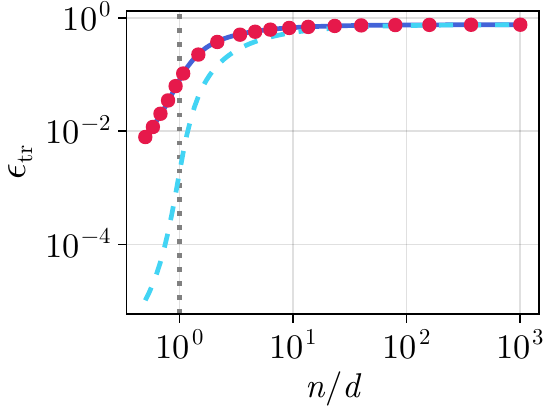}
    }\hfill
    \subfloat[YP MSD, $\lambda=10^{-4}$]{
      \includegraphics[width=0.23\textwidth]{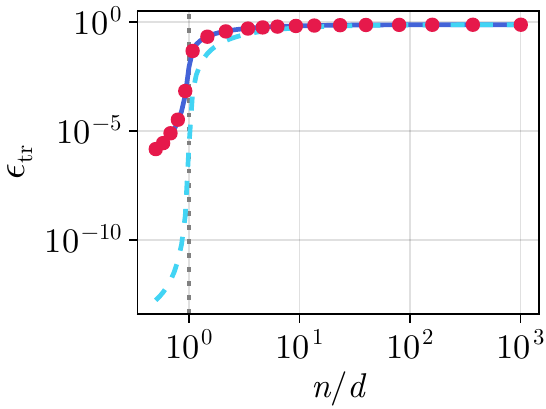}
    }
    \caption{Verification of variational Gaussian approximation (VGA) in linear regression. Both generalization and training errors are shown. (a)-(l): synthetic data. (m)-(p): real-world data.Markers represent true generalization and training errors evaluated by large data of size $n_0=10^4$. Lines are predictions of VGA.}
    \label{fig: linreg}
\end{figure*}

\subsubsection{Synthetic data: teacher–student scenario}
We first consider synthetic data in a teacher–student setting. 
Here the inputs are generated from a standard normal distribution, and the outputs $y$ are produced either by a linear model or by a two–layer neural network (2NN):
\begin{align}
  x &\sim \gN(0, I_d), \\
  y &= \begin{cases}
    \langle x, w_0\rangle + \zeta, & \text{linear}, \\
    \tfrac{1}{\sqrt{K}} \sum_{k=1}^K v_k \, \sigma(\langle x, w_{0,k}\rangle) + \zeta, & \text{2NN},
  \end{cases}
\end{align}
where $\sigma:\R\to\R$ denotes an activation function, $\zeta \sim \gN(0, \Delta)$ is measurement noise, and the true model parameters $w_0, v_k, w_{0,k}$ are independently drawn from normal distributions.  
Throughout the synthetic cases, we set $d=100$, $K=50$, and $\Delta=0.01$.  

Based on this generative model, we construct datasets of size $n_0=10^4$ and use them to approximate the averages over $z\sim\pdata$, evaluate the predicted generalization error, and compare with simulations.  Recall that this size $n_0$ can be different from $n$ that appears in stationarity condition.

Since the components of $x$ are i.i.d., we restrict ourselves to the simplified structure $q_w = \bar{q}_w I_d$ and $\chi_w = \bar{\chi}_w I_d$.  

Figure~\ref{fig: linreg} (a)–(l) show the comparison between the predictions of VGA and the experiments. In all cases, VGA provides a quantitatively accurate prediction.

\subsubsection{Real-world data}
\label{subsubsection: real-world data}

\begin{figure*}[t]
  \captionsetup[subfloat]{captionskip=-1truemm}
  \centering
    \subfloat[$(n_0,\lambda)=(515, 10^{-2})$]{
      \includegraphics[width=0.23\textwidth]{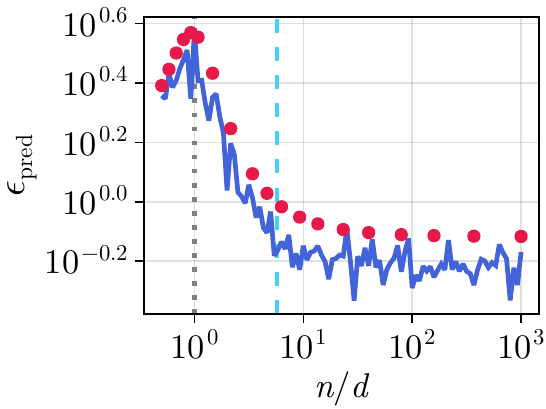}
    }\hfill
    \subfloat[$(n_0,\lambda)=(1031, 10^{-2})$]{
      \includegraphics[width=0.23\textwidth]{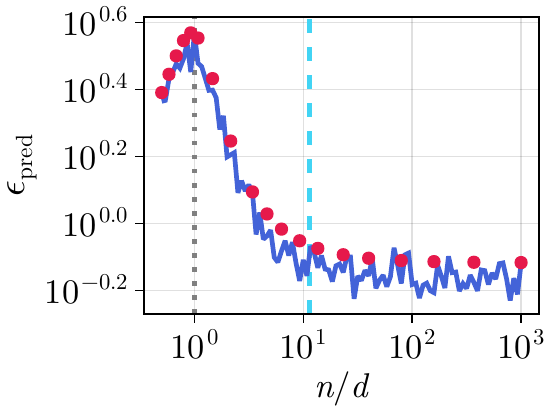}
    }\hfill
    \subfloat[$(n_0,\lambda)=(2061, 10^{-2})$]{
      \includegraphics[width=0.23\textwidth]{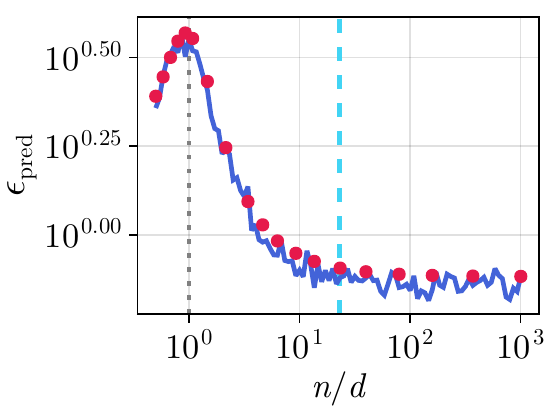}
    }\hfill
    \subfloat[$(n_0,\lambda)=(5153, 10^{-2})$]{
      \includegraphics[width=0.23\textwidth]{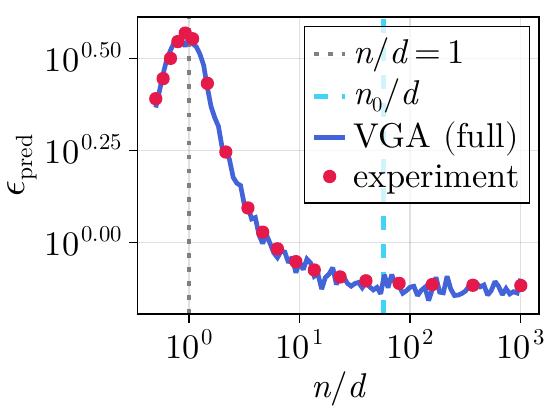}
    }\\
    \subfloat[$(n_0,\lambda)=(515, 10^{-4})$]{
      \includegraphics[width=0.23\textwidth]{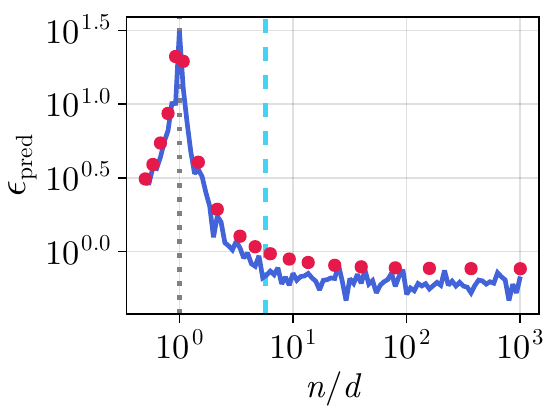}
    }\hfill
    \subfloat[$(n_0,\lambda)=(1031, 10^{-4})$]{
      \includegraphics[width=0.23\textwidth]{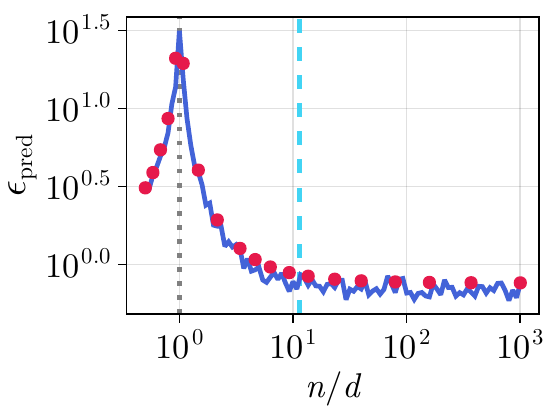}
    }\hfill
    \subfloat[$(n_0,\lambda)=(2061, 10^{-4})$]{
      \includegraphics[width=0.23\textwidth]{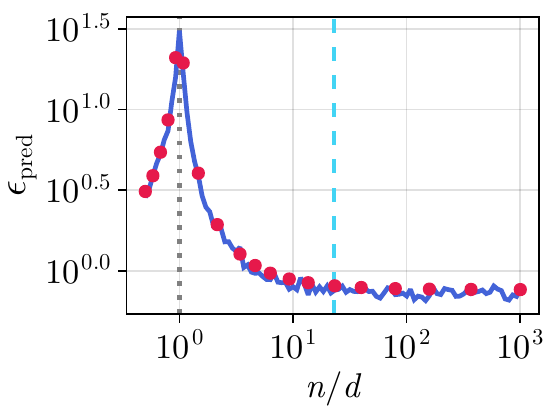}
    }\hfill
    \subfloat[$(n_0,\lambda)=(5153, 10^{-4})$]{
      \includegraphics[width=0.23\textwidth]{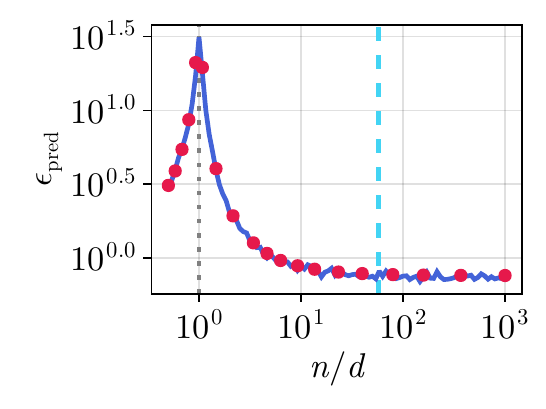}
    }
    \\
    \subfloat[$(n_0,\lambda)=(515, 10^{-2})$]{
      \includegraphics[width=0.23\textwidth]{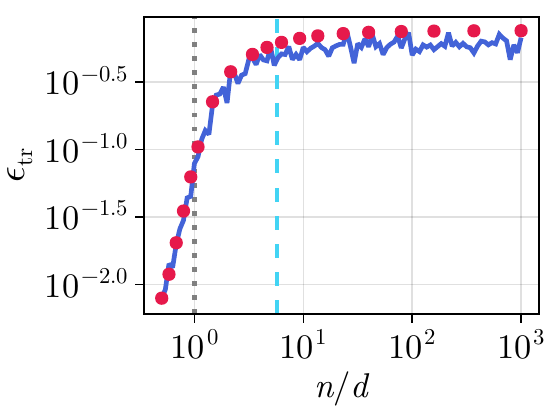}
    }\hfill
    \subfloat[$(n_0,\lambda)=(1031, 10^{-2})$]{
      \includegraphics[width=0.23\textwidth]{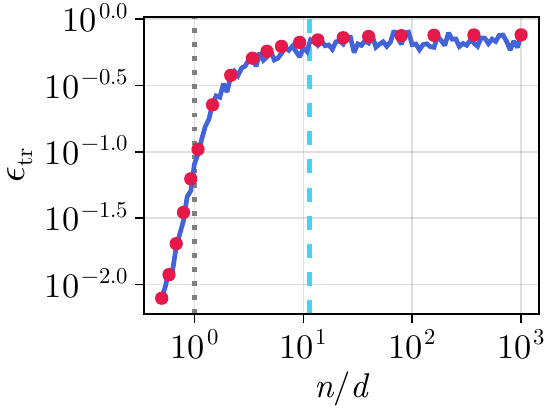}
    }\hfill
    \subfloat[$(n_0,\lambda)=(2061, 10^{-2})$]{
      \includegraphics[width=0.23\textwidth]{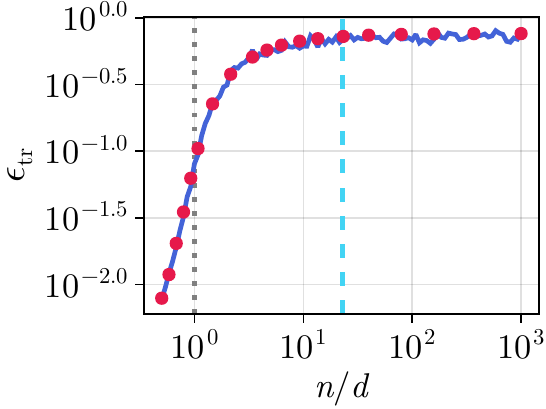}
    }\hfill
    \subfloat[$(n_0,\lambda)=(5153, 10^{-2})$]{
      \includegraphics[width=0.23\textwidth]{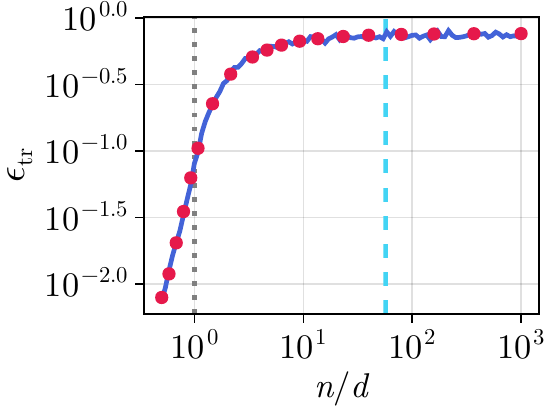}
    }\\
    \subfloat[$(n_0,\lambda)=(515, 10^{-4})$]{
      \includegraphics[width=0.23\textwidth]{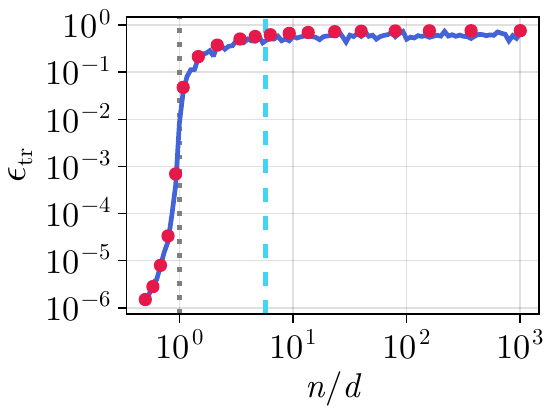}
    }\hfill
    \subfloat[$(n_0,\lambda)=(1031, 10^{-4})$]{
      \includegraphics[width=0.23\textwidth]{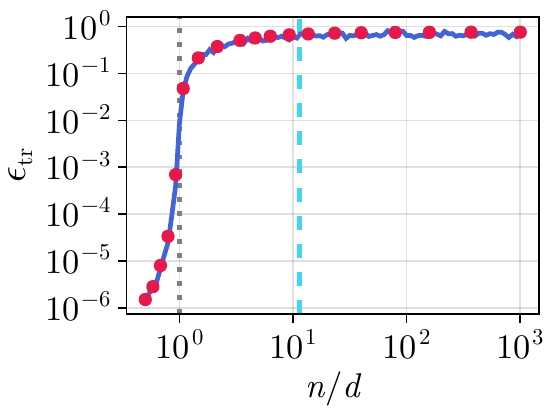}
    }\hfill
    \subfloat[$(n_0,\lambda)=(2061, 10^{-4})$]{
      \includegraphics[width=0.23\textwidth]{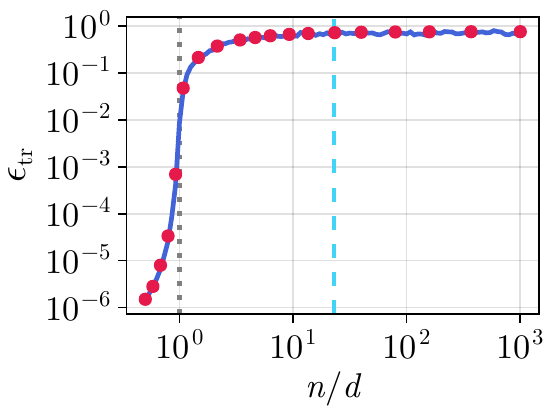}
    }\hfill
    \subfloat[$(n_0,\lambda)=(5153, 10^{-4})$]{
      \includegraphics[width=0.23\textwidth]{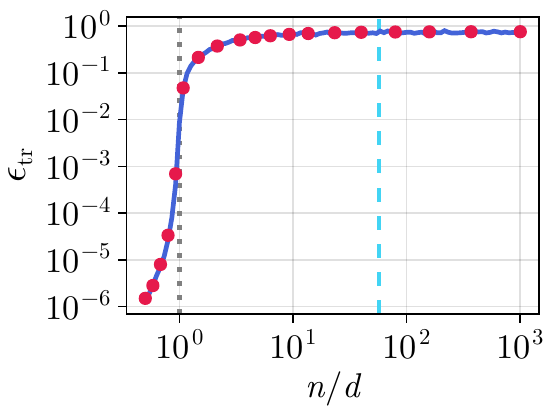}
    }
    \caption{The dependence of Learning curves for the YP MSD dataset on $n_0$. (a)-(h): generalization error $\bar{\epsilon}_{\mathrm{pred}}$. (i)-(p): training error $\bar{\epsilon}_{\rm tr}$. The average $\E_{z\sim \pdata}$ is approximated using a dataset of size $n_0$. Blue lines: VGA predictions. Red circles: experiments. Cyan dashed line: $n_0/d$.}
    \label{fig: size dep}
\end{figure*}

We next consider real-world data, using the Year Prediction MSD (YPMSD) dataset \cite{year_prediction_msd_203}. 
This dataset consists of the task of predicting the release year of a song from 90-dimensional audio features, with a total of $515{,}345$ samples. All features are standardized to have zero mean and unit variance.

For the choice of $n_0$, we first use 20\% of the data to approximate $\E_z$ ($n_0\simeq 10^5$), and the remaining 80\% to estimate the true generalization error. Later, the dependence of the predictions on $n_0$ will be discussed.  For the structure of $\chi_w$ and $q_w$, we consider two settings: one restricted to diagonal matrices and the other allowing full matrices.

The results for $n_0\simeq10^5$ are shown in Figure~\ref{fig: linreg}(m)-(p).  
In all cases VGA yields quantitatively accurate predictions when full matrices are used, while the diagonal approximation exhibits deviations for small sample sizes.  
This indicates that correlations among features play a crucial role in this dataset.
These results demonstrate that the validity of the variational approximation extends beyond controlled synthetic settings to real-world datasets.

Next, we check the dependence of the predictions on $n_0$.
Figure~\ref{fig: size dep} illustrates the dependence of the learning curves on $n_0$. 
Here, we varied $n_0$ from $515$ to $5153$, and for each estimate of the learning result with data size $n$, a different subset of size $n_0$ was sampled from the overall dataset to approximate the average over $z$. Also, for $q_w$ and $\chi_w$, only full matrices are considered.
For $n_0\simeq 500$, the estimates deviate from the true values and have large fluctuations, while for $n_0\simeq 2000$, the estimates become almost exact. It should be noted that accurate predictions are obtained even for $n \gg n_0$. In this sense, our approach is essentially different from bootstrap analyses 
\cite{10.1214/aos/1176344552,efron1994introduction}, 
where the variance vanishes when extrapolation to large data sizes is considered, even though the core idea behind the derivation of the GC formalism is similar to the bootstrap method.

\subsection{Single-basis ansatz}
Until now, we have considered a general vector $m_w$ as the direction in which the parameter symmetry is broken.  
However, in practice one may have prior knowledge about this direction.  
Within the present variational framework, such prior knowledge can naturally be incorporated by restricting the structure of the trial Hamiltonian.  

As the simplest example, which is particularly relevant in teacher–student scenarios or in the classification of two-component Gaussian mixtures, we consider the case where the symmetry-breaking direction is known, but its amplitude is not. Concretely, we assume 
\begin{equation}
  m_w \propto \bar{m}_w w_0, \qquad \bar{m}_w \in \mathbb{R},
\end{equation}
with a fixed reference vector $w_0$.  
The stationarity condition for $\bar{m}_w$ then takes the form
\begin{equation}
  \bar{m}_w = \frac{
    \frac{n}{d}\,\mathbb{E}_{z\sim\pdata}\left[\frac{\langle x,w_0 \rangle\,(y-m_f(x))}{1+\chi_f(x)} \right]
  }{
    \lambda + \frac{n}{d}\,\mathbb{E}_{z\sim\pdata}\left[\frac{\langle x,w_0 \rangle^2}{1+\chi_f(x)} \right]
  }.
\end{equation}
This expression makes it clear that $\bar{m}_w$ is determined by the (properly normalized) correlation between the residual $y-m_f(x)$ and the projection of the input $x$ onto the direction $w_0$.

In particular, consider the teacher–student model 
\[
  y = \langle x, w_0 \rangle + \zeta, 
  \qquad \zeta \sim \mathcal{N}(0,\Delta),
\]
and suppose that in the limit $d\to\infty$ the susceptibility $\chi_f(x)$ concentrates to a scalar $\bar{\chi}_w$, independent of $x$.  
In this case,
\begin{equation}
  \bar{m}_w = \frac{\frac{n/d}{1+\bar{\chi}_w}}{\lambda + \frac{n/d}{1+\bar{\chi}_w}}.
\end{equation}
This reproduces the well-known expression in solvable teacher–student scenario.

Although solvable settings yield compact closed forms, the variational approximation has the advantage of keeping the explicit dependence on data in the stationarity condition.  This provides a clearer view of which aspects of the data determine the parameter values.

\section{Summary and Conclusion}
\label{section: summary}
In this work, we have developed a replica analysis of inference and learning in parametric models.  
Unlike conventional approaches that focus on solvable teacher–student scenarios in the thermodynamic limit, we considered inference under general data distributions and finite system sizes.  

To this end, we introduced a grand canonical formalism that replaces dataset averages by virtual sampling from an infinitely large data reservoir, and applied a variational approximation to the resulting replicated system.  
Within this framework, the stationarity conditions for the variational parameters was derived without performing an analytic average over the data, allowing them to be determined adaptively for each given data distribution.  
A key technical point of our approach is that it changes the standard assumption in replica analyses: instead of requiring that the average over data can be carried out analytically, our method only requires that the trial Hamiltonian is sufficiently simple.
This perspective revealed connections to well-known information criteria in statistics and machine learning.  
As a concrete application, we analyzed linear regression and demonstrated that the method can yield learning curves even for problems involving real-world datasets, where the data-generating distribution is not known explicitly.
Although the basic idea of GC formalism was already given by Malzahn and Opper \cite{malzahn2002statistical,NIPS2000_ed519dac,NIPS2001_26f5bd4a, Malzahn_2005}, we believe the present work provides a clearer view for analysis in parameter space.

Natural extensions of this work include analyses of more complex models, such as multilayer neural networks, for which exact solutions have not been obtained yet except for limited scenarios.  Although the analysis of neural networks would become highly involved even with simple trial Hamiltonians, promising directions include structured scenarios such as sparse teacher–student models \cite{yoshino2020complex, yoshino2023spatially}. Also, in recent years, non-parametric variational approximation methods with neural networks have advanced significantly. Hence, the development of numerical variational methods would be also an important direction. Furthermore, hypothesis testing methods based on replica analysis have often relied on state evolution equations of solvable settings \cite{javanmard2014hypothesis, takahashi2018statistical, sur2019modern, na2023compressed}, and our approach may relate to directions aiming to propose methods with a more direct connection to observables \cite{bellec2025observable}.

Finally, although we have focused on the standard variational inference for approximate replica analysis, we briefly comment on other approaches. In principle, one could also consider methods such the Plefka expansion \cite{plefka1982convergence} and the expectation propagation / expectation-consistent approximation \cite{10.5555/2074022.2074067,PhysRevLett.86.3695,PhysRevE.64.056131,JMLR:v6:opper05a}, which are frequently used in spin-glass literature. These methods are closedly related to the optimization problem $\min_{\tilde{p}}\KL(p||\tilde{p})$, where $\tilde{p}$ denotes the approximate distribution and $p$ the target distribution, in contrast to the present formulation that is based on $\min_{\tilde{p}} \KL(\tilde{p}||p)$. Consequently, the objective function itself is different from Eq.\eqrefflat{eq: variational free energy}, and it would be more natural to employ the canonical formalism \eqrefflat{eq: replicated system canonical} rather than the GC formalism as in the studies by Yoshino \cite{yoshino2020complex,yoshino2023spatially}. From the viewpoint of approximation accuracy, it is unclear to the author which objective is better in general. In any case, approximate analysis of replicated system with general data distribution is at an early stage, and exploring various approximation schemes will be important.

\begin{acknowledgments}
This work was supported by JSPS KAKENHI Grant Numbers 22H05117 and 23K16960, and JST ACT-X Grant Number JPMJAX24CG.
\end{acknowledgments}

\appendix

\section{Grand canonical}
\label{appendix: grand canonical}

In subsection~\ref{subsection: grand canonical}, the average over datasets $\Dn$ was represented by drawing samples with replacement from a sufficiently large dataset $\tilde{D} = \{\tilde{z}_i\}_{i=1}^{\tilde{n}}$ with $\tilde{n} \gg n$, and then approximate this procedure by independent Poisson sampling. 
In this construction, the total number of sampled data points $\sum_{i=1}^{\tilde{n}} \tilde{c}_i$, with $\tilde{c}_i \sim {\rm Poisson}(n/\tilde{n})$, fluctuates around the mean $n$, with standard deviation $\sqrt{n}$, which is negligible compared to $n$ itself.
Thus, considering the GC formulation corresponds to a situation in which the sample size itself also fluctuates according to a Poisson distribution with mean $n$, in addition to the randomness in drawing each data point from $\pdata$.

Including this fluctuation, the replicated system may be written as
\begin{align}
  \frac{
    \sum_{k=0}^\infty \frac{e^{-n} n^k}{k!} 
    \E_{z\sim\pdata}\left[ e^{-\beta \sum_{a=1}^r l(\theta^a;z)} \right]^k
    e^{-\frac{\beta\lambda}{2}\sum_{a=1}^r\|w^a\|_2^2}
  }{
    \sum_{k=0}^\infty \frac{e^{-n} n^k}{k!} \Xi_r^{\beta,k}
  }\nonumber 
  \\
  \propto
  \exp\!\left(
    n \, \E_{z\sim\pdata}\left[
      e^{-\beta \sum_{a=1}^r l(\theta^a;z)}
    \right]
    -\frac{\beta\lambda}{2}\sum_{a=1}^r\|w^a\|_2^2
  \right).
\end{align}
This expression coincides with the GC formalism \eqrefflat{eq: replicated system grand canonical}. 

The representation of the normalization constant implies that in GC formalism, the partition function can be written as
\begin{equation}
    \Xi_{r,\gc}^{\beta,n} = \sum_{k=0}^\infty n^k \frac{1}{k!}\Xi_{r}^{\beta,k},
\end{equation}

This transform is similar to the one from canonical ensemble to grand canonical ensemble with fugacity $n$. In this sense, $\tilde{D}$ can be regarded as a reservoir. Also, it may be possible to interpret that the factor $k!$ reflect the label-exchange symmetry of the data points. The computation in this section is equivalent to the one in the main text. Nevertheless, we expect that it would be useful when considering more elaborate replica systems, such as those used in the replica analysis of bootstrap methods \cite{malzahn2003approximate, obuchi2019semi, 10206490, takahashi2024replica, takahashi2025replica}.

\section{Properties of the stationary point in a toy model}
\label{appendix: simplest case}

\begin{figure*}[t]
  \centering
    \centering
    \subfloat[$(-\partial_\chi \gF, -\partial_q\gF)$]{
      \includegraphics[width=0.44\textwidth]{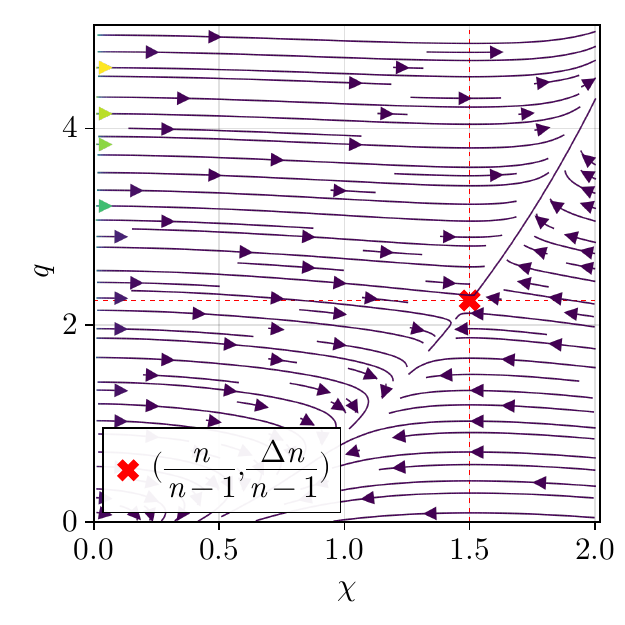}
    }\hfill
    \subfloat[$(-\partial_\chi \gF, +\partial_q\gF)$]{
      \includegraphics[width=0.44\textwidth]{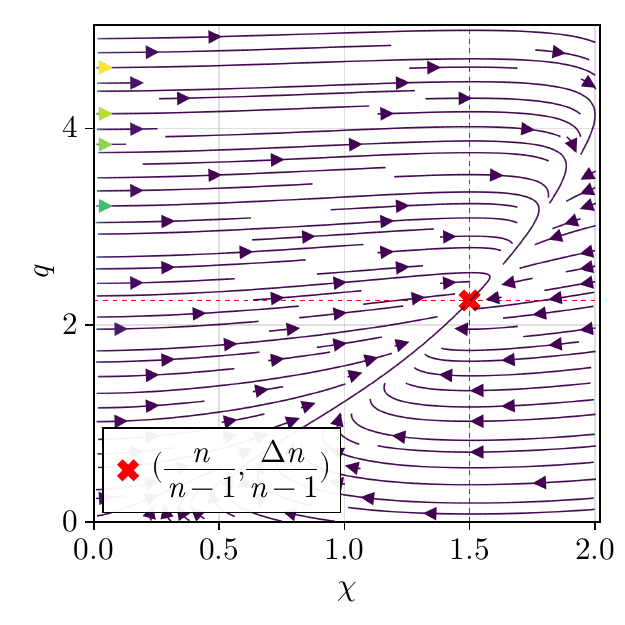}
    }\hfill
    \caption{Gradient fields of the gradients of the free energy $\gF \equiv \lim_{\beta\to\infty, r\to0}\gF_{r,\gc}^{\beta,n}$.  
    (a): the steepest descent direction of $\gF$.  
    (b): the case where the gradient with respect to $q$ is inverted.}
    \label{fig: stream free energy}
\end{figure*}

In this section, by considering a simple estimation problem, we demonstrate that the stationary conditions for the trial Hamiltonian do not, in general, correspond either to maximization or minimization of the leading term of the variational free energy, $\lim_{r\to0} r^{-1}\gF_{r,\gc}^{\beta,n}$.

Let us assume that the data consist of a one-dimensional random variable $z \in \R$, with the loss function $l(\theta;z) = (z-\theta)^2/2$, parameter $\theta \in \R$, and $\lambda=0$.  
For simplicity, we also assume $\E_{z\sim\pdata}[z] = 0$ and $\E_{z\sim\pdata}[z^2] = \Delta_z$.  
In the limit $\beta \to \infty$, the estimator $\hat{\theta}(\Dn)$ reduces to the sample mean $n^{-1}\sum_{i=1}^n z_i$.  
Of course, for $n \gg 1$, the central limit theorem guarantees that this converges to a Gaussian distribution with mean zero and variance $\Delta_z/n$ \cite{wasserman2013all}.  
Nevertheless, let us ignore this fact and attempt to analyze the fluctuations using the replica method.

In this case, decomposing $\hat{\theta}(\Dn) = \eta + \xi$ with $\eta \sim \gN(0, \chi/(\beta n))$ representing thermal fluctuations and $\xi \sim \gN(0, q/n)$ representing quenched randomness, we obtain
\begin{align}
  &\gF \equiv \lim_{\substack{r\to0 \\ \beta\to\infty}} \frac{1}{\beta r}\gF_{r,\gc}^{\beta,n} 
  = \frac{q}{\chi} - \frac{n}{1+\chi/n}(\Delta_z + \frac{q}{n}), \\
  &\frac{\chi^\ast}{n} = \frac{1}{n-1}, 
  \quad \frac{q^\ast}{n} = \frac{\Delta_z}{n-1}.
\end{align}
Thus, $\hat{\theta}(\Dn) \sim \gN(0, \sqrt{\Delta_z/(n-1)})$, showing that the expected behavior is indeed recovered.  
On the other hand, this stationary point is neither a minimum nor a maximum of $\gF$.  
Figure \ref{fig: stream free energy} shows the gradient fields $(-\partial_\chi \gF, -\partial_q \gF)$ and $(-\partial_\chi \gF, +\partial_q \gF)$.  
From the figure, we see that the stationary point corresponds to a minimum in the $\chi$ direction but a maximum in the $q$ direction.  
Therefore, unlike in standard variational approximations, one cannot determine the trial Hamiltonian parameters simply by maximization or minimization.  
Unfortunately, it remains an open question which variables should be maximized and which minimized in general.

\bibliography{main}

\end{document}